\documentclass[journal,draftcls,onecolumn,12pt,twoside]{IEEEtran}

\usepackage{cite}

\usepackage[cmex10]{amsmath}
\usepackage{mathtools}
\allowdisplaybreaks[4]

\usepackage{array}
\usepackage{mdwmath}
\usepackage{mdwtab}
\usepackage{hyperref}	
	
	\usepackage{bbm}
\usepackage{url}
\usepackage[normalem]{ulem}
	
\usepackage{graphicx} 
\usepackage{epsfig} 
\usepackage{epstopdf}
\usepackage{amssymb}  
\usepackage{verbatim}
\usepackage[usenames,dvipsnames,svgnames]{xcolor}
\usepackage{multirow} 
\usepackage{dsfont}
\graphicspath{ {./images/} }
\usepackage{verbatim}
\usepackage{wrapfig}

\def\lpr{\left\{}
\def\rpr{\right\}}

\newtheorem{theorem}{Theorem}

\newtheorem{lemma}{Lemma}

\newtheorem{proof}{Proof}
\newtheorem{assumption}{Assumption}

\newcommand{\mA}{\mathcal{A}}
\newcommand{\mX}{\mathcal{X}}

\newcommand{\mN}{[N]}
\newcommand{\cP}{\mathcal{P}}

\renewcommand{\hat}{\widehat}

\def\lpr{\left\{}
\def\rpr{\right\}}

\newcommand{\mE}{\mathbb{E}}	

\newcommand{\eq}[1]{\begin{align}#1\end{align}}
\newcommand{\seq}[1]{\begin{subequations}#1\end{subequations}}
\newcommand{\lb}[1]{\left\{ \begin{array}{ll} #1 \end{array} \right.}

\newcommand{\E}{\mathbb{E}}
 \newcommand{\nn}{\nonumber}
\newcommand{\cX}{\mathcal{X}}
\newcommand{\cZ}{\mathcal{Z}}
\newcommand{\cY}{\mathcal{Y}}
\newcommand{\cA}{\mathcal{A}}

\newcommand{\tsigma}{\tilde{\sigma}}
\newcommand{\hz}{\hat{z}}

\newcommand{\cH}{\mathcal{H}}
\newcommand{\tgamma}{\tilde{\gamma}}

\newcommand{\defeq}{\buildrel\triangle\over =}

\newcommand{\pushright}[1]{\ifmeasuring@ #1 \else\omit\hfill$\displaystyle#1$\fi\ignorespaces}
\newcommand{\pushleft}[1]{\ifmeasuring@ #1 \else\omit$\displaystyle#1$\hfill\fi\ignorespaces}

\begin{document}




\title{Markov perfect equilibria in non-stationary mean-field games}


\author{Deepanshu Vasal%
\thanks{University of Texas, Austin, {dvasal@utexas.edu}}
} 
\maketitle

\begin{abstract}
We consider both finite and infinite horizon discounted dynamic mean-field games where there is a large population of homogeneous players sequentially making strategic decisions and each player is affected by other players through an aggregate population state. Each player has a private type that only she observes and all players commonly observe a mean-field population state which represents the empirical distribution of other players' types.
Such games have been studied in the literature under simplifying assumption that population state dynamics are stationary. In this paper, we consider non-stationary population state dynamics and present a novel backward recursive algorithm to compute Markov perfect equilibrium (MPE) that depend on both, a player's private type, and current (dynamic) population state. Each step in this algorithm consists of solving a fixed-point equation. We provide conditions on model parameters for which there exists such an MPE. Using this algorithm, we study a security problem in cyber-physical system where infected nodes put negative externality on the system, and each node makes a decision to get vaccinated. We numerically compute MPE of the game.
\end{abstract}




\section{Introduction}
With increasing amount of integration of technology in our society and with recent advancements in computation and algorithmic technologies, there is an unprecedented scale of interaction among people and devices.
With technologies such as ride sharing platforms and social media apps completely integrated, and new technologies such as cyber physical systems, large scale renewable energy, electric vehicles, cryptocurrencies and smart grid on the horizon, there is paramount need to design and understand the behavior of such \emph{large scale} interactions and their impact on our society. In this paper, we present a new methodology to analyze such interactions through mean-field dynamic games.

Dynamic games is a powerful tool to model such sequential strategic interaction among selfish players, introduced by Shapley in~\cite{Sh53}. Discrete-time dynamic games with Markovian structure have been studied extensively to model many practical applications, in engineering as well as economics literature, such as dynamic auctions, security, markets, traffic routing, wireless systems, social learning, oligopolies-- i.e. competition among firms~\cite{BaOl98, FiVr12}.

In dynamic games with perfect and symmetric information, subgame perfect equilibrium (SPE) is an appropriate equilibrium concept and there exists a backward recursive algorithm to find all the SPEs of these games (refer to~\cite{OsRu94, FuTi91book, samuelson2006} for a more elaborate discussion). Maskin and Tirole in~\cite{MaTi01} introduced the concept of Markov perfect equilibrium (MPE) where players' strategies depend on a coarser Markovian state of the systems, instead of the whole history of the game which grows exponentially with time and thus becomes unwieldy. This is a refinement of the SPE. In general, there exists a backward recursive methodology to compute MPE of the game. Some prominent examples of the application of MPE include~\cite{ErPa95, BeVa96, AcRo01}. Ericson and Pakes in~\cite{ErPa95} model industry dynamics for firms' entry, exit and investment participation, through a dynamic game with symmetric information, compute its MPE, and prove ergodicity of the equilibrium process. Bergemann and V{\" a}lim{\"a}ki in~\cite{BeVa96} study a learning process in a dynamic oligopoly with strategic sellers and a single buyer, allowing for price competition among sellers. They study MPE of the game and its convergence behavior. Acemo\u{g}lu and Robinson in~\cite{AcRo01} develop a theory of political transitions in a country by modeling it as a repeated game between the elites and the poor, and study its MPE.

However, when the number of players is large, computing MPE becomes intractable.
To model the behavior of large population strategic interactions, mean-field games were introduced independently by Huang, Malham\'e, and Caines~\cite{HuMaCa06}, and Lasry and Lions~\cite{LaLi07}. In such games, there are large number of homogenous strategic players, where each player has infinitesimal affect on system dynamics and is affected by other players through a mean-field population state. There have been a number of applications such as economic growth, security in networks, oil production, volatility formation, population dynamics (see ~\cite{La08,GuLaLi11,SuMa19,HuMa16,HUMa17,HuMa17cdc, AdJoWe15} and references therein). 

To motivate our problem better, consider the following application. Consider a dynamic energy market, where in each period, a large number of different suppliers bid their estimated power outputs to an independent system operator (ISO) that formulates the market mechanism to  determine the prices assessed to the different suppliers. Each supplier wants to maximize its overall return, which depends on its cost of production of energy, which is its private information, and the market-determined prices which depend on all the bids. Each bidder is thus affected not by other individual bidders, but an aggregated population state of everybody else. 

In this paper, to model the scenarios described above we consider discounted infinite-horizon dynamic mean-field games where there is a large population of homogenous players each having a private type. Each player sequentially makes strategic decisions and is affected by other players through a mean-field population state. Each player has a private type that evolves through a controlled Markov process which only she observes and all players observe the current population state which is the distribution of other players' types. In such games, the mean-field state evolves through McKean Vlasov \emph{forward} equation given a policy of the players. And the equilibrium policy satisfies the Bellman \emph{backward} equation, given the the mean-field states. Thus to compute equilibrium, one needs to solve the coupled backward and forward fixed-point equation in the mean-field and the equilibrium policy.

In~\cite{AdJoWe15}, authors study \emph{stationary} equilibria of a mean-field game where they make simplifying assumption on the model that the players are \emph{oblivious} with respect to the mean-field statistics, and are playing in the limit such that the mean-field distribution has converged. This allows them to decouple the mean-field dynamics with that of the rest of the game.

In this paper, we consider a general model where players are \emph{cognizant} i.e. they actively observe the current population state (which need not have converged) and act based on that population state and their own private state.
We provide a novel backward recursive algorithm to compute \emph{non-stationary, signaling} Markov perfect equilibrium (MPE) of that game. We also provide sufficient conditions for existence of MPE for both finite and infinite-horizon. In general, this algorithm could be used to relook the applications studied in mean-field games, to study equilibria with non-stationary mean-field statistics.

Using this framework, we consider malware spread problem in a cyber-physical system where nodes get infected by an independent random process and for each node, there is a higher risk of getting infected due to negative externality imposed by other infected players. At each time $t$, each player privately observes its own state and publicly observes the population of infected nodes, based on which it has to make a decision to repair or not. Using our algorithm, we find equilibrium strategies of the players which are observed to be non-decreasing in the healthy population state. 

Our algorithm is motivated by recent developments in the theory of dynamic games with asymmetric information in~\cite{VaSiAn16arxiv, VaAn16allerton, VaAn16cdc, OuTaTe17,Ta17}, where authors in these works have considered different models of such games and provided a sequential decomposition framework to compute Markovian perfect Bayesian equilibria of such games. 

The paper is structured as follows. In Section~\ref{sec:Model}, we present model, notation and background. In section~\ref{sec:methodology}, we present our main results where we present algorithm to compute MPE for both finite and infinite horizon game, and also present existence results. We present a numerical example in Section~\ref{sec:example}. We conclude in Section~\ref{sec:Concl}.

\subsection{Notation}
We use uppercase letters for random variables and lowercase for their realizations. For any variable, subscripts represent time indices and superscripts represent player identities. We use notation $ -i$ to represent all players other than player $i$ i.e. $ -i = \{1,2, \ldots i-1, i+1, \ldots, N \}$. We use notation $a_{t:t'}$ to represent the vector $(a_t, a_{t+1}, \ldots a_{t'})$ when $t'\geq t$ or an empty vector if $t'< t$. We use $a_t^{-i}$ to mean $(a^1_t, a^2_{t}, \ldots, a_t^{i-1}, a_t^{i+1} \ldots, a^N_{t})$ . We remove superscripts or subscripts if we want to represent the whole vector, for example $a_t$  represents $(a_t^1, \ldots, a_t^N) $. We denote the indicator function of any set $A$ by $\mathbbm{1}\{A\}$. 
For any finite set $\mathcal{S}$, $\mathcal{P}(\mathcal{S})$ represents space of probability measures on $\mathcal{S}$ and $|\mathcal{S}|$ represents its cardinality. We denote by $P^{\sigma}$ (or $E^{\sigma}$) the probability measure generated by (or expectation with respect to) strategy profile $G$. We denote the set of real numbers by $\mathbb{R}$. For a probabilistic strategy profile of players $(\sigma_t^i)_{i\in [N]}$ where probability of action $a_t^i$ conditioned on $z_{1:t},x_{1:t}^i$ is given by $\sigma_t^i(a_t^i|z_{1:t},x_{1:t}^i)$, we use the short hand notation $\sigma_t^{-i}(a_t^{-i}|z_{1:t},x_{1:t}^{-i})$ to represent $\prod_{j\neq i} \sigma_t^j(a_t^j|z_{1:t},x_{1:t}^j)$.   
All equalities and inequalities involving random variables are to be interpreted in \emph{a.s.} sense.

\section{Model and Background}
\label{sec:Model}
We consider both finite and infinite-horizon discrete-time large population sequential game as follows. There are $N$ homogenous players, where $N$ tends to $\infty$. We denote the set of homogenous players by $[N]$ and with some abuse of notation, set of time by [T] for both finite and infinite time horizon. In each period $t\in[T]$, player $i\in[N]$ observes a private type $x_t^i\in\cX = \{1,2,\cdots, N_x\}$ and a common observation $y_t\in\cY$, takes action $a_t^i\in\cA = \{1,2,\cdots, N_a \}$, and receives a reward $R(x_t^i,a_t^i,y_t)$ which is a function of its current type $x_t^i$, action $a_t^i$ and the common observation $y_t$. The common observation $z_t = (z_t(1),z_t(2),\ldots,z_t(N_x))$ be the fraction of population having type $x\in\cX$ at time $t$ i.e.
\eq{
z_t(x) = \frac{1}{N}\sum_{i=1}^N \mathbbm{1}\{x_t^i = x\},
} 
where $\sum_{i=1}^{N_x} z_t(i) = 1$.
Player $i$'s type evolve as a controlled Markov process,
\eq{
x_{t+1}^i = f_x(x_t^i, a_t^i, z_t, w_t^i).
}
The random variables $(w_t^i)_{i,t}$ are assumed to be mutually independent across players and across time. We also write the above update of $x_t^i$ through a kernel, $x_{t+1}^i\sim Q_x(\cdot|x_t^i, a_t^i, z_t)$.

In any period $t$, player $i$ observes $(z_{1:t},x_{1:t}^i)$. She takes action $a_t^i$ according to a behavioral strategy $\sigma^i = (\sigma_t^i)_t$, where $\sigma_t^i:(\cZ)^{t}\times\mathcal{X}^t \to \mathcal{P}(\mathcal{A})$. We denote the space of such strategies as $\mathcal{K}^{\sigma}$. This implies $A_t^i\sim \sigma_t^i(\cdot|z_{1:t},x_{1:t}^i)$. We denote $\mathcal{H}_t^c = \mathcal{Z}^t$ to be the space of population states $z_{1:t}$ till time $t$. We denote $\mathcal{H}_t^i=\mathcal{Z}^t \times \mathcal{X}^t$ to be set of observed histories $(z_{1:t},x_{1:t}^i)$ of player $i$.

For finite time-horizon game, $\mathbb{G}_{T}$, each player wants to maximize its total expected discounted reward over a time horizon $T$, discounted by discount factor $0<\delta\leq1$, 
\eq{
J^{i,T} :=\E^{\sigma} \left[\sum_{t=1}^T \delta^{t-1} R(X_t^i,A_t^i,Z_t) \right].
} 

For the infinite time-horizon game, $\mathbb{G}_{\infty}$, each player wants to maximize its total expected discounted reward over an infinite-time horizon discounted by discount factor $0<\delta<1$, 
\eq{
J^{i,\infty} :=\E^{\sigma} \left[\sum_{t=1}^\infty \delta^{t-1} R(X_t^i,A_t^i,Z_t) \right].
} 

\subsection{Solution concept: MPE}
The Nash equilibrium (NE) of $\mathbb{G}_{T}$ is defined as strategies $\tsigma = (\tsigma_t^i)_{i\in[N],t\in[T]}$ that satisfy, for all $i\in[N]$, 
\eq{
\E^{(\tsigma^i,\tsigma^{-i})}[\sum_{t=1}^T \delta^{t-1} R(X_t^i,A_t^i,Z_t) ]\geq \E^{(\sigma^i,\tsigma^{-i})}[\sum_{t=1}^T \delta^{t-1} R(X_t^i,A_t^i,Z_t)],
} 
For sequential games, however, a more appropriate equilibrium concept is Markov perfect equilibrium (MPE)~\cite{MaTi01}, which we use in this paper. We note that an MPE is also a Nash equilibrium of the game, although not every Nash equilibrium is an MPE.
An MPE $(\tsigma)$ satisfies sequential rationality such that for $\mathbb{G}_T$, $\forall i\in[N], t \in [T], h^{i}_t \in \cH^i_t, {\sigma^{i}}$,

\eq{
&\E^{(\tsigma^{i} \tsigma^{-i})}[\sum_{n=t}^T \delta^{n-t} R(X_n^i,A_n^i,Z_n)|z_{1:t},x_{1:t}^i ] \geq \E^{({\sigma}^{i} \tsigma^{-i})}[\sum_{n=t}^T \delta^{n-t} R(X_n^i,A_n^i,Z_n)|z_{1:t},x_{1:t}^i ], \;\; \;\;   \label{eq:seqeq2}
} 
NE and MPE for $\mathbb{G}_{\infty}$ are defined in a similar way where summation in the above equations is taken such that $T$ is replaced by $\infty$.

\section{A methodology to compute MPE }
\label{sec:methodology}
In this section, we will provide a backward recursive methodology to compute MPE for both $\mathbb{G}_{T}$ and $\mathbb{G}_{\infty}$.
We will consider Markovian equilibrium strategies of player $i$ which depend on the common information at time $t$, $z_{t}$, and on its current type $x_t^i$.\footnote{Note however, that the unilateral deviations of the player are considered in the space of all strategies.}  Equivalently, player $i$ takes action of the form $A_t^i\sim \sigma_t^i(\cdot|z_t,x_t^i)$. Similar to the common agent approach in~\cite{NaMaTe13}, an alternate and equivalent way of defining the strategies of the players is as follows. We first generate partial function $\gamma_t^i:\cX\to\cP(\cA)$ as a function of $z_t$ through an equilibrium generating function $\theta_t^i:\cZ\to(\cX\to\cP(\cA))$ such that $\gamma_t^i = \theta_t^i[z_t]$. Then action $A_t^i$ is generated by applying this prescription function $\gamma_t^i$ on player $i$'s current private information $x_t^i$, i.e. $A_t^i\sim \gamma_t^i(\cdot|x_t^i)$. Thus $A_t^i\sim \sigma_t^i(\cdot|z_{t},x_t^i) = \theta_t^i[z_t](\cdot|x_t^i)$.

We are only interested in symmetric equilibria of such games such that $A_t^i\sim \gamma_t(\cdot|x_t^i) = \theta_t[z_t](\cdot|x_t^i)$ i.e. there is no dependence of $i$ on the strategies of the players.

For a given symmetric prescription function $\gamma_t = \theta[z_t]$, the statistical mean-field $z_t$ evolves according to the discrete-time McKean Vlasov equation, $\forall y\in\cX$:
\eq{
z_{t+1}(y) =\sum_{x\in\cX}\sum_{a\in \cA} z_t(x)\gamma_t(a|x)Q_x(y|x,a,z_t), \label{eq:z_update}
}
which implies
\eq{
z_{t+1}= \phi(z_t,\gamma_t).
}

\subsection{Backward recursive algorithm for $\mathbb{G}_{T}$} \label{sec:fhbr}
In this subsection, we will provide a methodology to generate symmetric MPE of $\mathbb{G}_{T}$ of the form described above.
We define an equilibrium generating function $(\theta_t)_{t\in[T]}$, where $\theta_t:\cZ\to\{\cX\to\mathcal{P}(\cA) \}$, where for each $z_t $, we generate $\tgamma_t = \theta_t[z_t]$. In addition, we generate a reward-to-go function $(V_t)_{t\in[T]}$, where $V_t:\cZ\times\cX\to\mathbb{R}$.
These quantities are generated through a fixed-point equation as follows.
\begin{itemize}
\item[1.] Initialize $\forall z_{T+1}, x_{T+1}^i\in \cX$,
\eq{
V_{T+1}(z_{T+1},x_{T+1}^i) \defeq 0.   \label{eq:VT+1}
}

\item[2.] For $t = T,T-1, \ldots 1, \ \forall z_t$, let $\theta_t[z_t] $ be generated as follows. Set $\tilde{\gamma}_t = \theta_t[z_t]$, where $\tilde{\gamma}_t$ is the solution of the following fixed-point equation\footnote{We discuss the existence of solution of this fixed-point equation in Section~\ref{sec:exists}}, $\forall i \in [N],x_t^i\in \cX$,
  \eq{
 \tilde{\gamma}_t(\cdot|x_t^i) \in  \arg\max_{\gamma_t(\cdot|x_t^i)} \E^{\gamma_t(\cdot|x_t^i)} \left[ R(x_t^i,A_t^i,z_t) +\delta V_{t+1}(\phi(z_t,\tgamma_t), X_{t+1}^{i}) | z_t,x_t^i\right] , \label{eq:m_FP}
  }
 where expectation in \eqref{eq:m_FP} is with respect to random variable $(A_t^i,X_{t+1}^{i})$ through the measure
$\gamma_t(a_t^i|x_t^i)Q_x(x_{t+1}^{i}|x_t^{i},a_t^i,z_t)$.
We note that the solution of~\eqref{eq:m_FP}, $\tgamma_t$, appears both on the left of~\eqref{eq:m_FP} and on the right side in the update of $z_t$, and is thus unlike the fixed-point equation found in Bayesian Nash equilibrium.

Furthermore, using the quantity $\tilde{\gamma}_t$ found above, define
\eq{
V_{t}(z_t,x_t^i) \defeq & \E^{\tilde{\gamma}_t(\cdot|x^i)} \left[ R(x_t^i,A_t^i,z_t) +\delta V_{t+1}(\phi(z_t,\tgamma_t), X_{t+1}^{i}) | z_t,x_t^{i}\right].  \label{eq:Vdef}
}
   \end{itemize}
   
Then, an equilibrium strategy is defined as 
\eq{
\tilde{\sigma}_t^i(a_t^i|z_{1:t},x_{1:t}^i) = \tilde{\gamma}_t(a_t^i|x_t^i), \label{eq:sigma_fh}
} 
where $\tilde{\gamma}_t = \theta[z_t]$. 

In the following theorem, we show that the strategy thus constructed is an MPE of the game.

\begin{theorem}
\label{Thm:Main}
A strategy $(\tsigma)$ constructed from the above algorithm is an MPE of the game i.e. $\forall t, h^{i}_t \in \cH^i_t, {\sigma^{i}}$,

\eq{
&\E^{(\tsigma^{i} \tsigma^{-i})}[\sum_{n=t}^T \delta^{n-t} R(X_n^i,A_n^i,Z_n)|z_{1:t},x_{1:t}^i ] \geq \E^{({\sigma}^{i} \tsigma^{-i})}[\sum_{n=t}^T \delta^{n-t} R(X_n^i,A_n^i,Z_n)|z_{1:t},x_{1:t}^i ], \;\; \;\;   \label{eq:prop}
} 
\end{theorem}
\proof
Please see Appendix~\ref{app:A}.
\endproof

\subsection{Converse}
In the following, we show that every Markovian mean field equilibria can be found using the above backward recursion.

\begin{theorem}[Converse]
	\label{thm:2}
	Let $\tsigma$ be a Markovian MPE of the mean field game. Then there exists an equilibrium generating function $\theta$
 that satisfies \eqref{eq:m_FP} in backward recursion $ \forall\ z_{1:t}$ such that  $\tsigma$ is defined using $\phi$.
\end{theorem}
\proof
	Please see Appendix~\ref{app:C}.
\endproof

\subsection{Backward recursive algorithm for $\mathbb{G}_{\infty}$} \label{sec:fhbr}
In this section, we consider the infinite-horizon problem $\mathbb{G}_{\infty}$, for which we assume the reward function $R$ to be absolutely bounded.

We define an equilibrium generating function $\theta:\cZ\to\{\cX\to\mathcal{P}(\cA) \}$, where for each $z_t $, we generate $\tgamma_t = \theta[z_t]$. In addition, we generate a reward-to-go function $V:\cZ\times\cX\to\mathbb{R}$.
These quantities are generated through a fixed-point equation as follows.

For all $z,$ set $\tilde{\gamma} = \theta[z]$. Then $(\tilde{\gamma},V)$ are solution of the following fixed-point equation\footnote{We discuss the existence of solution of this fixed-point equation in Section~\ref{sec:exists}}, $\forall z\in\cZ,x^i\in \cX$,
\seq{
\label{eq:m_FP_ih}
  \eq{
 \tilde{\gamma}(\cdot|x^i) &\in \arg\max_{\gamma(\cdot|x^i)} \E^{\gamma(\cdot|x^i)} \left[ R(x^i,A^i,z) +\delta V(\phi(z,\tgamma), X^{i'}) | z,x^i\right] , \\
 V(z,x^i) &=\  \E^{\tilde{\gamma}(\cdot|x^i)} \left[ R(x^i,A^i,z) +\delta V (\phi(z,\tgamma), X^{i'}) | z,x^{i}\right]. 
  }
  }
 where expectation in \eqref{eq:m_FP_ih} is with respect to random variable $(A^i,X^{i,\prime})$ through the measure
$\gamma(a^i|x^i)Q_x(x^{i'}|x^{i},a^i,z)$.

Then an equilibrium strategy is defined as 
\eq{
\tilde{\sigma}^i(a_t^i|z_{1:t},x_{1:t}^i) = \tilde{\gamma}(a_t^i|x_t^i), \label{eq:sigma_ih}
}
where $\tilde{\gamma} = \theta[z_t]$.

The following theorem shows that the strategy thus constructed is an MPE of the game.

\begin{theorem}
\label{thih}
A strategy $(\tsigma)$ constructed from the above algorithm is an MPE of the game i.e. $\forall t, h^{i}_t \in \cH^i_t, {\sigma^{i}}$,

\eq{
&\E^{(\tsigma^{i} \tsigma^{-i})}[\sum_{n=t}^\infty \delta^{n-t} R(X_n^i,A_n^i,Z_n)|z_{1:t},x_{1:t}^i ] \geq \E^{({\sigma}^{i} \tsigma^{-i})}[\sum_{n=t}^\infty \delta^{n-t} R(X_n^i,A_n^i,Z_n)|z_{1:t},x_{1:t}^i ], \;\; \;\;   \label{eq:prop_ih}
} 
\end{theorem}
\proof
Please see Appendix~\ref{app:D}.
\endproof

\subsection{Converse}
In the following, we show that every Markovian mean field equilibria can be found using the above backward recursion.
\begin{theorem}[Converse]
	\label{thm:2ih}
	Let $\tsigma$ be a Markovian MPE of the mean field game. Then there exists an equilibrium generating function $\theta$
 that satisfies \eqref{eq:m_FP} in backward recursion $ \forall\ z_{1:t}$ such that  $\tsigma$ is defined using $\theta$.
\end{theorem}
\proof
	Please see Appendix~\ref{app:idih}.
\endproof

\section{Existence}
\label{sec:exists}
In this section, we discuss sufficient conditions for the existence of a solution of the fixed-point equations~\eqref{eq:m_FP} and~\eqref{eq:m_FP_ih}. 

\begin{assumption}[A1]
Let the reward function $R(x_t,a_t,z_t)$ and the state update kernel $Q_x(x_{t+1}|x_t,a_t,z_t)$ be continuous functions in $z_t$.
\end{assumption}

We note that the above equation implies that the reward function is bounded.

\begin{theorem}
Under assumption~(A1), there exists solution of the fixed-point equations~\eqref{eq:m_FP} and~\eqref{eq:m_FP_ih} for every $t$.
\end{theorem}
\proof
Under the assumption (A1), it has been shown in~\cite{DoGaGa19} that there exists Markovian MPE of both the finite and infinite horizon game. Theorem~\ref{thm:2} and Theorem~\ref{thm:2ih} show that all Markovian MPE can be found using backward recursion for the finite and infinite horizon problems respectively. This proves that under (A1), there exists a solution of~\eqref{eq:m_FP} and~\eqref{eq:m_FP_ih} for every $t$.
\endproof

\section{Numerical Example: Cyber physical security}
\label{sec:example}

We consider a security problem in a cyber physical network with positive externalities. It is discretized version of the malware problem presented in~\cite{HuMa16,HUMa17,HuMa17cdc,JiAnWa11}. Some other applications of this model include flu vaccination, entry and exit of firms, investment, network effects. In this model, suppose there are large number of cyber physical nodes where each node has a private state $x_t^i\in\{0,1\}$ where $x_t^i= 0$ represent `healthy' state and $x^i_t= 1$ is the infected state. Each node can take action $a_t^i\in\{0,1\}$, where $a_t^i= 0$ implies ``do nothing" and $a_t^i=1$ implies repair. The dynamics are given by
\eq{
x_{t+1}^i = \lb{x_t^i + (1-x_t^i)w_t^i\;\; \text{ for } a_t^i = 0\\
0 \hspace{63pt}\;\;\text{ for } a_t^i = 1.
}
}
where $w_t^i\in \{0,1\}$ is a binary valued random variable with $P(w_t^i = 1) = q$, which represents the probability of a node getting infected. Thus if a node doesn't do anything, it could get infected with certain probability, however, if it takes repair action, it comes back to the healthy state. 
Each node gets a reward 
\eq{
r(x^i_t,a^i_t,z_t) =-(k+z_t(1))x^i_t-\lambda a_t^i.
}
where $z_t(1)$ is the mean-field population state being 1 at time $t$, $\lambda$ is the cost of repair and $(k+z_t(1))$ represents the risk of being infected.
We pose it as an infinite horizon discounted dynamic game.
We consider parameters $k= 0.2,\lambda= 0.5, \delta = 0.9, q=0.9$ for numerical results presented in Figures~1-4.

\begin{figure}[htbp] 
   \centering
   \includegraphics[width=3.5in]{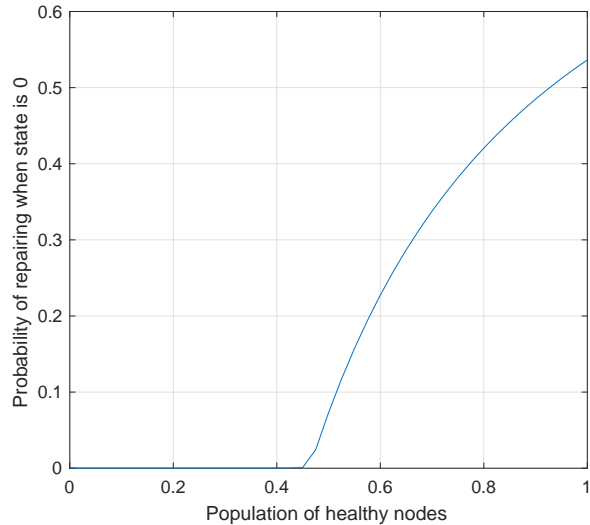} 
   \caption{$\gamma(1|0)$: Probability of choosing action 1, given $x^i = 0 $}
   \label{fig:example}
\end{figure}
\begin{figure}[htbp] 
   \centering

      \includegraphics[width=3.5in]{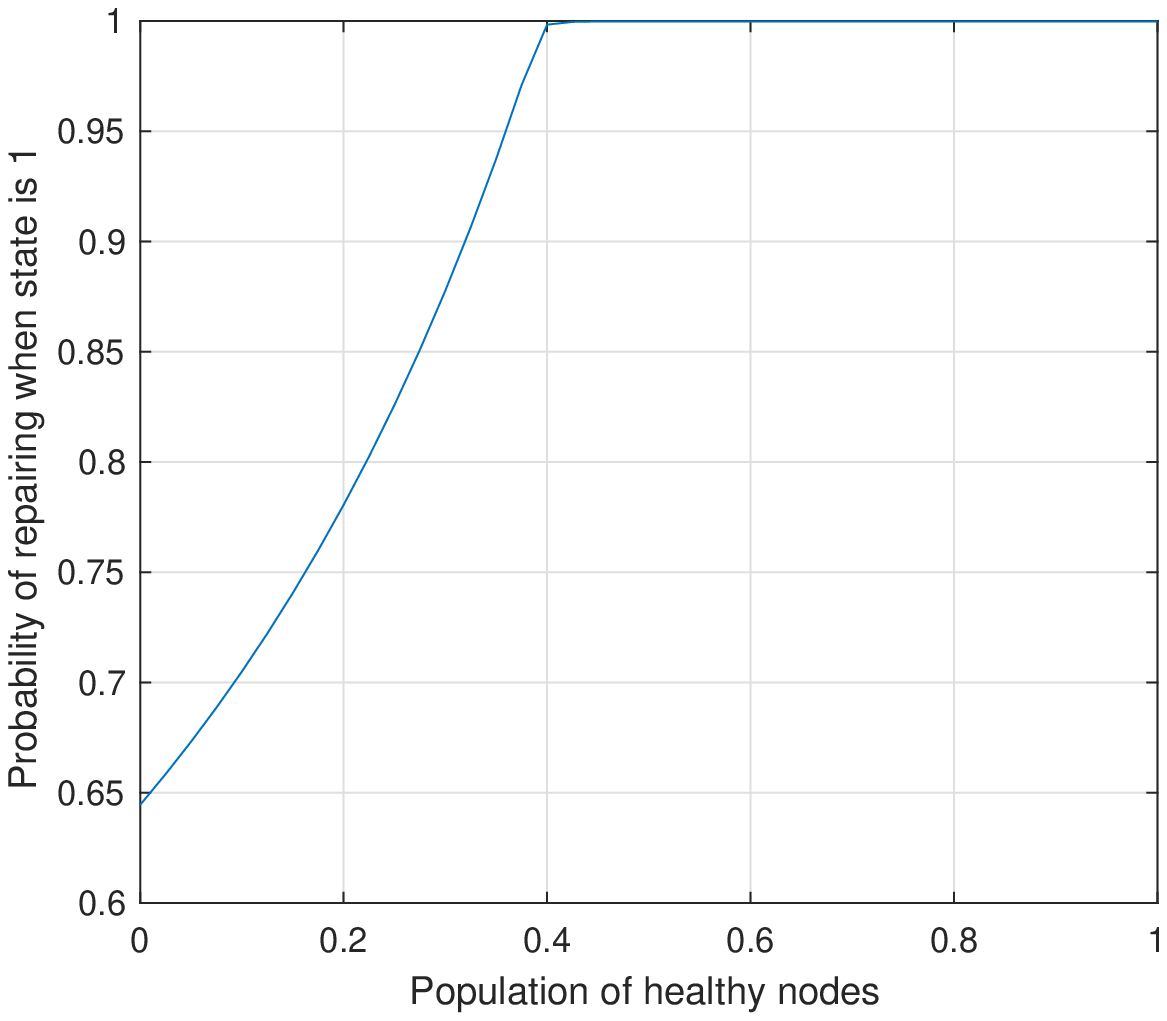} 
   \caption{$\gamma(1|1)$: Probability of choosing action 1, given $x^i = 1 $}
   \label{fig:example}
\end{figure}
\begin{figure}[htbp] 
   \centering
   \includegraphics[width=3.5in]{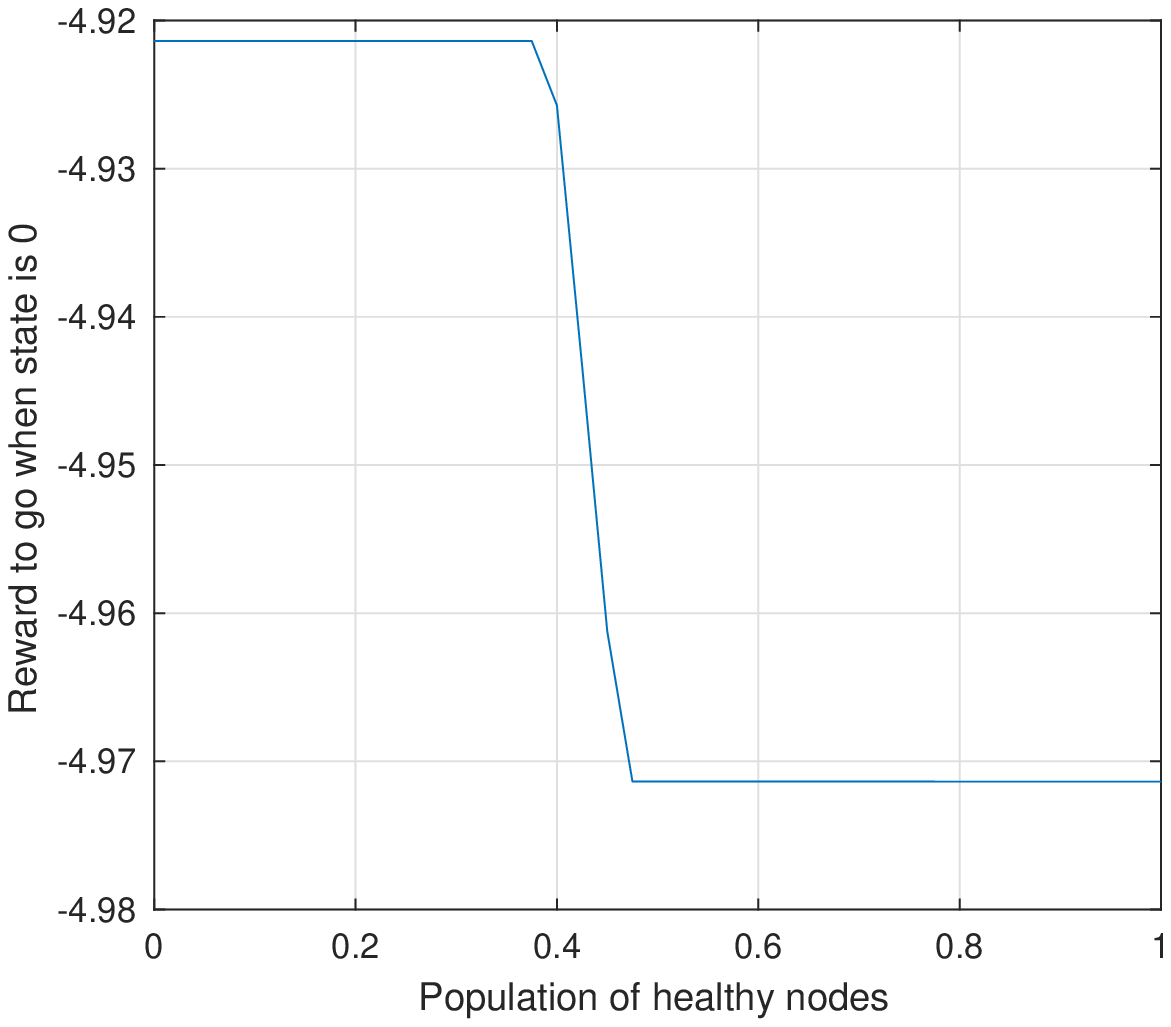} 
   \caption{$V(g(0),0)$ : Reward to go when state $x^i=0$.}
   \label{fig:example}
\end{figure}

\begin{figure}[htbp] 
   \centering
     \includegraphics[width=3.5in]{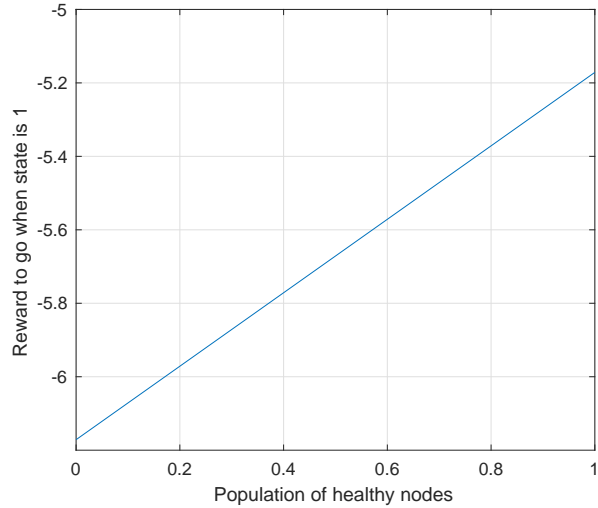} 
   \caption{$V(g(0),1)$ : Reward to go when state $x^i=1 $}
   \label{fig:example}
\end{figure}



%

\section{Conclusion}
\label{sec:Concl}
In this paper, we consider both finite and infinite horizon, large population dynamic game where each player is affected by others through a mean-field population state. We present a novel backward recursive algorithm to compute non-stationary, signaling Markov perfect equilibria (MPE) for such games, where each player's strategy depends on its current private type and current mean-field population state. The non-triviality in the problem is that the update of population state is coupled to the strategies of the game, and is managed in the algorithm through unique construction of the fixed-point equations~\eqref{eq:m_FP},\eqref{eq:m_FP_ih}. We proved the existence of such equilibrium. Using this algorithm, we considered a malware propagation problem where we numerically computed equilibrium strategies of the players. In general, this algorithm could instrumental in studying non-stationary equilibria in a number of applications such as financial markets, social learning, renewable energy.

\section*{Acknowledgments}

The author would like to acknowledge the support of Simons Grant \#26-7523-99 and Department of Defense grant \#W911NF1510225. The author thanks Francois Baccelli and Sriram Vishwanath for encouragement and support.


%


%
%
%
 \appendices
 \section{}
\label{app:A}
\proof
We prove~\eqref{eq:prop} using induction and the results in Lemma~\ref{lemma:2}, and \ref{lemma:1} proved in Appendix~\ref{app:B}.
\seq{
For base case at $t=T$, $\forall i\in [N], (z_{1:T}, x_{1:T}^i)\in \mathcal{H}_{T}^i, \sigma^i$
\eq{
\E^{\tsigma_{T}^{i} \tsigma_{T}^{-i} }\left\{  R(X_T^i,A_T^i,Z_T) \big\lvert z_{1:T}, x_{1:T}^i \right\}
&=
V_T(z_T, x_T^i)  \label{eq:T2a}\\
&\geq \E^{\sigma_{T}^{i} \tsigma_{T}^{-i}} \left\{ R(X_T^i,A_T^i,Z_T) \big\lvert z_{1:T}, x_{1:T}^i \right\},  \label{eq:T2}
}
}
where \eqref{eq:T2a} follows from Lemma~\ref{lemma:1} and \eqref{eq:T2} follows from Lemma~\ref{lemma:2} in Appendix~\ref{app:B}.

Let the induction hypothesis be that for $t+1$, $\forall i\in [N], z_{1:t+1} \in (\mathcal{H}_{t+1}^c), x_{1:t+1}^i \in (\cX)^{t+1}, \sigma^i$,
\seq{
\eq{
 \E^{\tsigma_{t+1:T}^{i} \tsigma_{t+1:T}^{-i}} \left\{ \sum_{n=t+1}^T \delta^{n-t-1}R(X_n^i,A_n^i,Z_n) \big\lvert z_{1:t+1}, x_{1:t+1}^i \right\} \\
 \geq
  \E^{\sigma_{t+1:T}^{i} \tsigma_{t+1:T}^{-i}} \left\{ \sum_{n=t+1}^T \delta^{n-t-1} R(X_n^i,A_n^i,Z_n) \big\lvert  z_{1:t+1}, x_{1:t+1}^i \right\}. \label{eq:PropIndHyp}
}
}
\seq{
Then $\forall i\in [N], (z_{1:t}, x_{1:t}^i) \in \mathcal{H}_{t}^i, \sigma^i$, we have
\eq{
&\E^{\tsigma_{t:T}^{i} \tsigma_{t:T}^{-i}} \left\{ \sum_{n=t}^T \delta^{n-t-1}R(X_n^i,A_n^i,Z_n) \big\lvert z_{1:t}, x_{1:t}^i \right\} \nonumber \\
&= V_t(z_{t}, x_t^i)\label{eq:T1}\\
&\geq \E^{\sigma_t^i \tsigma_t^{-i}} \left\{ R(X_t^i,A_t^i,Z_t) + \delta V_{t+1}^i (Z_{t+1}, X_{t+1}^i) \big\lvert z_{1:t}, x_{1:t}^i \right\}  \label{eq:T3}\\
&= \E^{\sigma_t^i \tsigma_t^{-i}} \left\{ R(X_t^i,A_t^i,Z_t) + \delta \E^{\tsigma_{t+1:T}^{i} \tsigma_{t+1:T}^{-i}} \left\{ \sum_{n=t+1}^T \delta^{n-t-1}R(X_n^i,A_n^i,Z_n) \big\lvert  z_{1:t},Z_{t+1}, x_{1:t}^i,X_{t+1}^i \right\}  \big\vert z_{1:t}, x_{1:t}^i \right\}  \label{eq:T3b}\\
&\geq \E^{\sigma_t^i \tsigma_t^{-i}} \left\{ R(X_t^i,A_t^i,Z_t) + \delta\E^{\sigma_{t+1:T}^{i} \tsigma_{t+1:T}^{-i} } \left\{ \sum_{n=t+1}^T \delta^{n-t-1}R(X_n^i,A_n^i,Z_n) \big\lvert z_{1:t},Z_{t+1}, x_{1:t}^i,X_{t+1}^i\right\} \big\vert z_{1:t}, x_{1:t}^i \right\}  \label{eq:T4} \\
&= \E^{\sigma_t^i \tsigma_t^{-i}} \big\{ R(X_t^i,A_t^i,Z_t) +  \delta\E^{\sigma_{t:T}^{i} \tsigma_{t:T}^{-i} } 
\left\{ \sum_{n=t+1}^T \delta^{n-t-1}R(X_n^i,A_n^i,Z_n) \big\lvert z_{1:t},Z_{t+1}, x_{1:t}^i,X_{t+1}^i\right\} \big\vert z_{1:t}, x_{1:t}^i \big\}  
\label{eq:T5}\\
&=\E^{\sigma_{t:T}^{i} \tsigma_{t:T}^{-i}} \left\{ \sum_{n=t}^T \delta^{n-t}R(X_n^i,A_n^i,Z_n) \big\lvert z_{1:t},  x_{1:t}^i \right\}  \label{eq:T6},
}
}
where \eqref{eq:T1} follows from Lemma~\ref{lemma:1}, \eqref{eq:T3} follows from Lemma~\ref{lemma:2}, \eqref{eq:T3b} follows from Lemma~\ref{lemma:1}, \eqref{eq:T4} follows from induction hypothesis in \eqref{eq:PropIndHyp} and \eqref{eq:T5} follows since the random variables involved in the right conditional expectation do not depend on strategies $\sigma_t^i$.
\endproof

\section{}
\label{app:B}
\begin{lemma}
\label{lemma:2}
$\forall t\in [T], i\in [N], (z_{1:t}, x_{1:t}^i)\in \mathcal{H}_t^i, \sigma^i_t$
\eq{
V_t(z_t, x_t^i) \geq \E^{\sigma_t^i \tsigma_t^{-i}} \left\{ R(X_t^i,A_t^i,Z_t) + \delta V_{t+1} (Z_{t+1}, X_{t+1}^i) \big\lvert  z_{1:t}, x_{1:t}^i \right\}.\label{eq:lemma2}
}
\end{lemma}

\proof
We prove this lemma by contradiction.

 Suppose the claim is not true for $t$. This implies $\exists i, \hat{\sigma}_t^i, \hat{z}_{1:t}, \hat{x}_{1:t}^i$ such that
\eq{
\E^{\hat{\sigma}_t^i \tsigma_t^{-i}} \left\{ R(X_t^i,A_t^i,Z_t) +  \delta V_{t+1} (Z_{t+1}, X_{t+1}^i) \big\lvert \hat{z}_{1:t},\hat{x}_{1:t}^i \right\} 
> V_t(\hz_t, \hat{x}_{t}^i).\label{eq:E8}
}
We will show that this leads to a contradiction.
Construct 
\begin{equation}
\hat{\gamma}^i_t(a_t^i|x_t^i) = \lb{\hat{\sigma}_t^i(a_t^i|\hat{z}_{1:t},\hat{x}_{1:t}^i) \;\;\;\;\; x_t^i = \hat{x}_t^i \\ \text{arbitrary} \;\;\;\;\;\;\;\;\;\;\;\;\;\; \text{otherwise.}  }
\end{equation}

Then for $\hat{z}_{1:t}, \hat{x}_{1:t}^i$, we have
\seq{
\eq{
&V_t(\hz_t, \hat{x}_t^i) 
\nn \\
&= \max_{\gamma_t(\cdot|\hat{x}_t^i)} \E^{\gamma_t(\cdot|\hat{x}_t^i) \tsigma_t^{-i}} \left\{ R(\hat{x}_t^i,A_t^i,\hz_t) + \delta V_{t+1} (\phi(\hz_t,\tgamma_t), X_{t+1}^i) \big\lvert \hz_t, \hat{x}_{t}^i \right\}, \label{eq:E11}\\
&\geq\E^{\hat{\gamma}_t^i(\cdot|\hat{x}_t^i) \tsigma_t^{-i}} \left\{ R(x_t^i,A_t^i,z_t) + \delta V_{t+1} (\phi(\hz_t,\tgamma_t), {X}_{t+1}^i) \big\lvert \hz_t,\hat{x}_{t}^i \right\}   
\\ \nn 
&=\sum_{a_t^i,x_{t+1}^i}   \left\{ R(\hat{x}_t^i,a_t^i,\hz_t) + \delta V_{t+1} (\phi(\hz_t,\tgamma_t), x_{t+1}^i)\right\}
\hat{\gamma}_t(a^i_t|\hat{x}_t^i)Q_t^i(x_{t+1}^i|\hat{x}_t^i,a_t^i,\hz_t)  
\\ \nn 
&= \sum_{a_t^i,x_{t+1}^i}  \left\{ R(\hat{x}_t^i,a_t^i,\hz_t) + \delta V_{t+1} (\phi(\hz_t,\tgamma_t), x_{t+1}^i)\right\}
\hat{\sigma}_t(a_t^i|\hat{z}_{1:t} ,\hat{x}_{1:t}^i) Q_t^i(x_{t+1}^i|\hat{x}_t^i,a_t^i,\hz_t) \label{eq:E9}\\
&= \E^{\hat{\sigma}_t^i \tsigma_t^{-i}} \left\{ R(\hat{x}_t^i,a_t^i,\hz_t)+ \delta V_{t+1} (\phi(\hz_t,\tgamma_t), X_{t+1}^i) \big\lvert \hat{z}_{1:t},  \hat{x}_{1:t}^i \right\}  \\
&> V_t(\hz_t, \hat{x}_{t}^i), \label{eq:E10}
}
where \eqref{eq:E11} follows from definition of $V_t$ in \eqref{eq:Vdef}, \eqref{eq:E9} follows from definition of $\hat{\gamma}_t^i$ and \eqref{eq:E10} follows from \eqref{eq:E8}. However this leads to a contradiction.
}
\endproof

\begin{lemma}
\label{lemma:1}
$\forall i\in [N], t\in [T], (z_{1:t}, x_{1:t}^i)\in \mathcal{H}_t^i$,
\begin{gather}
V_t(z_{t}, x_t^i) =
\E^{\tsigma_{t:T}^{i} \tsigma_{t:T}^{-i}} \left\{ \sum_{n=t}^T \delta^{n-t}R(X_n^i,A_n^i,Z_n) \big\lvert  z_{1:t}, x_{1:t}^i \right\} .
\end{gather} 
\end{lemma}

\proof
%
\seq{
We prove the lemma by induction. For $t=T$,
\eq{
 \E^{\tsigma_{T}^{i} \tsigma_{T}^{-i} } \left\{  R(X_T^i,A_T^i,Z_T) \big\lvert z_{1:T},  x_{1:T}^i \right\}
 &= \sum_{a_T^i} R(x_T^i,a_T^i,z_T) \tsigma_{T}^{i}(a_T^i|z_{T},x_{T}^i) \\
 &= V_T(z_{T}, x_T^i) \label{eq:C1},
}
}
where \eqref{eq:C1} follows from the definition of $V_t$ in \eqref{eq:Vdef}.
Suppose the claim is true for $t+1$, i.e., $\forall i\in [N], t\in [T], (z_{1:t+1}, x_{1:t+1}^i)\in \mathcal{H}_{t+1}^i$
\begin{gather}
V_{t+1}(z_{t+1}, x_{t+1}^i) = \E^{\tsigma_{t+1:T}^{i} \sigma_{t+1:T}^{-i}}
\left\{ \sum_{n=t+1}^T \delta^{n-t-1}R(X_n^i,A_n^i,Z_n) \big\lvert z_{1:t+1}, x_{1:t+1}^i \right\} 
\label{eq:CIndHyp}.
\end{gather}
Then $\forall i\in [N], t\in [T], (z_{1:t}, x_{1:t}^i)\in \mathcal{H}_t^i$, we have
\seq{
\eq{
&\E^{\tsigma_{t:T}^{i} \tsigma_{t:T}^{-i} } \left\{ \sum_{n=t}^T \delta^{n-t} R(X_n^i,A_n^i,Z_n) \big\lvert  z_{1:t}, x_{1:t}^i \right\} 
\nonumber 
\\
&=  \E^{\tsigma_{t:T}^{i} \tsigma_{t:T}^{-i} } \left\{R(X_t^i,A_t^i,Z_t)  \right.
\nonumber \\ 
&\left.+\delta \E^{\tsigma_{t:T}^{i} \tsigma_{t:T}^{-i} }  \left\{ \sum_{n=t+1}^T \delta^{n-t-1}R(X_n^i,A_n^i,Z_n)\big\lvert z_{1:t},  Z_{t+1}, x_{1:t}^i,X_{t+1}^i\right\} \big\lvert z_{1:t},  x_{1:t}^i \right\} \label{eq:C2}
\\
&=  \E^{\tsigma_{t:T}^{i} \tsigma_{t:T}^{-i} } \left\{R(X_t^i,A_t^i,Z_t) \right.
\nonumber 
\\
&\left.  +\delta\E^{\tsigma_{t+1:T}^{i} \tsigma_{t+1:T}^{-i} }\left\{ \sum_{n=t+1}^T \delta^{n-t-1}R(X_n^i,A_n^i,Z_n)\big\lvert z_{1:t},Z_{t+1}, x_{1:t}^i,X_{t+1}^i\right\} \big\lvert z_{1:t}, x_{1:t}^i \right\} \label{eq:C3}
\\
&=  \E^{\tsigma_{t:T}^{i} \tsigma_{t:T}^{-i} } \left\{R(X_t^i,A_t^i,Z_t) +  \delta V_{t+1}(Z_{t+1}, X_{t+1}^i) \big\lvert  z_{1:t}, x_{1:t}^i \right\} 
\label{eq:C4}
\\
&=  \E^{\tsigma_{T}^{i} \tsigma_{T}^{-i}} \left\{R(X_t^i,A_t^i,Z_t) +  \delta V_{t+1}(Z_{t+1}, X_{t+1}^i) \big\lvert  z_{1:t}, x_{1:t}^i \right\} 
\label{eq:C5}
\\
&=V_{t}(z_t, x_t^i) \label{eq:C6},
}
}
\eqref{eq:C4} follows from the induction hypothesis in \eqref{eq:CIndHyp}, \eqref{eq:C5} follows because the random variables involved in expectation, $X_t^i,A_t^i,Z_t,Z_{t+1},X_{t+1}^i$ do not depend on $\tsigma_{t+1:T}^{i} \sigma_{t+1:T}^{-i}$ and \eqref{eq:C6} follows from the definition of $V_t$ in \eqref{eq:Vdef}.
\endproof

\section{}
\label{app:C}
\proof
We prove this by contradiction. Suppose for any equilibrium generating function $\theta$ that generates an MPE $\tsigma$, there exists $t\in[T], i\in[N], z_{1:t}\in\cH_t^c,$ such that \eqref{eq:m_FP} is not satisfied for $\theta$
i.e. for $\tgamma_t = \theta_t[z_t] = \tsigma_t(\cdot|z_t,\cdot)$,
\eq{
 \tilde{\gamma}_t \not\in \arg\max_{\gamma_t(\cdot|x_t^i)} \E^{\gamma_t(\cdot|x_t)} \left\{ R_t(X_t^i,A_t^i,Z_t) + V_{t+1}(\phi(Z_t,\tilde{\gamma}_t), X_{t+1}^i) \big\lvert  x_t^i,z_t \right\} . \label{eq:FP4}
  }
  Let $t$ be the first instance in the backward recursion when this happens. This implies $\exists\ \hat{\gamma}_t$ such that
  \eq{
  \E^{\hat{\gamma}_t(\cdot|x_t^i)} \left\{ R_t(X_t^i,A_t^i,Z_t)+ V_{t+1} (\phi(Z_t, \tilde{\gamma}_t), X_{t+1}^i) \big\lvert  z_{1:t},x_{1:t}^i\right\}
  \nn\\
  > \E^{\tgamma_t(\cdot|x_t^i)} \left\{ R_t(X_t^i,A_t^i) + V_{t+1} (\phi(Z_t, \tilde{\gamma}_t), X_{t+1}^i) \big\lvert  z_{1:t},x_{1:t}^i \right\} \label{eq:E1}
  }
  This implies for $\hat{\sigma}_t(\cdot|z_t,\cdot) = \hat{\gamma}_t$,
  \eq{
  &\E^{\tsigma_{t:T}} \left\{ \sum_{n=t}^T R_n(X_n^i,A_n^i,Z_n) \big\lvert  z_{1:t-1}, x_{1:t}^i \right\}
  \nn\\
\\
  &= \E^{\tsigma_t^{i},\tsigma_t^{-i}} \left\{ R_t(X_t^i,A_t^i,Z_t) + \E^{\tsigma_{t+1:T}^{i} \tsigma_{t+1:T}^{-i}}   \left\{ \sum_{n=t+1}^T R_n(X_n^i,A_n,Z_n) \big\lvert z_{1:t-1},Z_{t+1}, x_{1:t}^i,X_{t+1}^i \right\}  \big\vert z_{1:t}, x_{1:t}^i \right\} \label{eq:E2}
  \\
  &=\E^{\tgamma_t(\cdot|x_t) \tilde{\gamma}^{-i}_t} \left\{ R_t(X_t^i,A_t^i,Z_t) + V_{t+1} (\phi(Z_t, \tilde{\gamma}_t), X_{t+1}^i) \big\lvert  z_t,x_t^i \right\} \label{eq:E3}
  \\
  &< \E^{\hat{\sigma}_t(\cdot|z_t,x_t^i) \tilde{\gamma}^{-i}_t} \left\{ R_t(X_t^i,A_t^i) + V_{t+1} (\phi(Z_t, \tilde{\gamma}_t), X_{t+1}^i) \big\lvert  z_t,x_t \right\}\label{eq:E4}
  \\
  &= \E^{\hat{\sigma}_t \tsigma_t^{-i}} \left\{ R_t(X_t^i,A_t^i,Z_t) +  \E^{\tsigma_{t+1:T}^{i} \tsigma_{t+1:T}^{-i}}\left\{ \sum_{n=t+1}^T R_n(X_n^i,A_n^i,Z_n) \big\lvert z_{1:t},Z_{t+1}, x_{1:t}^i,X_{t+1}^i\right\} \big\vert z_{1:t}, x_{1:t} ^i\right\}\label{eq:E5}
  \\
  &=\E^{\hat{\sigma}_t,\tsigma_{t+1:T}^{i} \tsigma_{t:T}^{-i}} \left\{ \sum_{n=t}^T R_n(X_n^i,A_n,Z_n) \big\lvert  z_{1:t}, x_{1:t}^i\right\},\label{eq:E6}
  }
  where \eqref{eq:E3} follows from the definitions of $\tgamma_t$ and Lemma~\ref{lemma:1}, \eqref{eq:E4} follows from \eqref{eq:E1} and the definition of $\hat{\sigma}_t$, \eqref{eq:E5} follows from Lemma~\ref{lemma:2}. However, this leads to a contradiction since $\tsigma$ is an MPE of the game.
\endproof

\section{}
\label{app:D}
	We divide the proof into two parts: first we show that the value function $ V $ is at least as big as any reward-to-go function; secondly we show that under the strategy $ \tsigma $, reward-to-go is $ V $. Note that $h_t^i := (z_{1:t},x_{1:t}^i)$.

\paragraph*{Part 1}
For any $ i \in \mN $, $ \sigma^i $ define the following reward-to-go functions
\begin{subequations} \label{eqihr2g}
\eq{
	W_t^{\sigma^i}(h_t^i) &= \mE^{\sigma^i,\tsigma^{-i}} \lpr \sum_{n=t}^\infty \delta^{n-t} R(X_n^i,A_n^i,Z_n) \mid h_t^i \rpr\\
	W_t^{\sigma^i,T}(h_t^i) &= \mE^{\sigma^i,\tsigma^{-i}} \lpr \sum_{n=t}^T \delta^{n-t} R(X_n^i,A_n^i,Z_n)
	+  \delta^{T+1-t} V(Z_{T+1},X^i_{T+1}) \mid h_t^i \rpr.
}
\end{subequations}
Since $ \mX,\mA $ are finite sets the reward $ R$ is absolutely bounded, the reward-to-go $ W_t^{\sigma^i}(h_t^i) $ is finite $ \forall $ $ i,t,\sigma^i,h_t^i $.

For any $ i \in \mN $, $ h_t^i \in \mathcal{H}_t^i $,
\eq{ \label{eqdc}
V\big(z_t,x_t^i\big) - W_t^{\sigma^i}(h_t^i)
= \Big[ V\big(z_t,x_t^i\big) - W_t^{\sigma^i,T}(h_t^i) \Big]
+ \Big[ W_t^{\sigma^i,T}(h_t^i) - W_t^{\sigma^i}(h_t^i) \Big]
	}
Combining results from Lemmas~\ref{thmfh2} and~\ref{lemfhtoih} in Appendix~\ref{app:D}, %
the term in the first bracket in RHS of~\eqref{eqdc} is non-negative. Using~\eqref{eqihr2g}, the term in the second bracket is
\begin{gather}\label{eqdiff}	
\left( \delta^{T+1-t} \right) \mE^{\sigma^i,\tsigma^{-i}} \Big\{- \sum_{n=T+1}^\infty \delta^{n-(T+1)} R(X_n^i,A_n^i,Z_n)
+ V(Z_{T+1},X^i_{T+1}) \mid h_t^i \Big\}.
\end{gather} 	
The summation in the expression above is bounded by a convergent geometric series. Also, $ V $ is bounded. Hence the above quantity can be made arbitrarily small by choosing $ T $ appropriately large. Since the LHS of~\eqref{eqdc} does not depend on $ T $, which implies,
\begin{gather}
V\big(z_t,x_t^i\big) \ge W_t^{\sigma^i}(h_t^i).
\end{gather}

\paragraph*{Part 2}
Since the strategy the equilibrium strategy $ \tsigma $ generated in~\eqref{eq:sigma_ih} is such that $\tsigma^{i}_t $ depends on $ h_t^i $ only through $ z_t $ and $ x_t^i $, the reward-to-go $ W_t^{\tsigma^{i}} $, at strategy $ \tsigma $, can be written (with abuse of notation) as
\begin{gather}
W_t^{\tsigma^{i}}(h_t^i) = W_t^{\tsigma^{i}}(z_t,x_t^i)
= \mE^{\tsigma} \lpr \sum_{n=t}^\infty \delta^{n-t} R(X_n^i,A_n^i,Z_n) \mid z_t,x_t^i \rpr.
\end{gather}

For any $ h_t^i \in \mathcal{H}_t^i $,
\begin{subequations}
\eq{
	W_t^{\tsigma^{i}}(z_t,x_t^i)
	&= \mE^{\tsigma} \lpr R(X_t^i,A_t^i,Z_t)	+ \delta W_{t+1}^{\tsigma^{i}}
	\big(\phi(z_t,\theta[z_t])),X_{t+1}^i\big)  \mid z_t,x_t^i \rpr\\
	V(z_t,x_t^i)
	&= \mE^{\tsigma} \Big\{ R(X_t^i,A_t^i,Z_t) + \delta V
	\big(\phi(z_t,\theta[z_t])),X_{t+1}^i\big)  \mid z_t,x_t^i \Big\}.
}
\end{subequations}
Repeated application of the above for the first $ n $ time periods gives
\begin{subequations}
\eq{
	W_t^{\tsigma^{i}}(z_t,x_t^i)
	&= \mE^{\tsigma}\Bigg\{ \sum_{m=t}^{t+n-1} \delta^{m-t} R(X_t^i,A_t^i,Z_t)
	+ \delta^{n}  W_{t+n}^{\tsigma^{i}}\big(Z_{t+n},X_{t+n}^i\big)  \mid z_t,x_t^i \Bigg\}
\\
	V(z_t,x_t^i)
	&= \mE^{\tsigma} \Bigg\{ \sum_{m=t}^{t+n-1} \delta^{m-t} R(X_t^i,A_t^i,Z_t)
	+ \delta^{n}  V\big(Z_{t+n},X_{t+n}^i\big)  \mid z_t,x_t^i \Bigg\}.
}
\end{subequations}
Taking differences results in
	\eq{
	W_t^{\tsigma^{i}}(z_t,x_t^i)  - V(z_t,x_t^i)
	=\delta^n \mE^{\tsigma}
	  \lpr W_{t+n}^{\tsigma^{i}}\big(Z_{t+n},X_{t+n}^i\big)
	- V\big(Z_{t+n},X_{t+n}^i\big) \mid z_t,x_t^i \rpr.
	}
Taking absolute value of both sides then using Jensen's inequality for $ f(x) = \vert x \vert $ and finally taking supremum over $ h_t^i $ reduces to
\eq{
\sup_{h_t^i} \big\vert W_t^{\tsigma^{i}}(z_t,x_t^i)  - V(z_t,x_t^i) \big\vert 
\le \delta^n \sup_{h_t^i}  \mE^{\tsigma}
 \lpr\big\vert W_{t+n}^{\tsigma^{i}}(Z_{t+n},X_{t+n}^i)
 - V(Z_{t+n},X_{t+n}^i) \big\vert  \mid z_t,x_t^i \rpr.
}
Now using the fact that $ W_{t+n},V$ are bounded and that we can choose $ n $ arbitrarily large, we get $ \sup_{h_t^i} \vert W_t^{\tsigma^{i}}(z_t,x_t^i)  - V(z_t,x_t^i) \vert = 0 $. 	

\section{}
\label{app:E}
In this section, we present three lemmas. Lemma~\ref{thmfh1} is intermediate technical results needed in the proof of Lemma~\ref{thmfh2}. Then the results in Lemma~\ref{thmfh2} and~\ref{lemfhtoih} are used in Appendix~\ref{app:C} for the proof of Theorem~\ref{thih}. The proof for Lemma~\ref{thmfh1} below isn't stated as it analogous to the proof of Lemma~\ref{lemma:2} from Appendix~\ref{app:B}, used in the proof of Theorem~\ref{Thm:Main} (the only difference being a non-zero terminal reward in the finite-horizon model).

Define the reward-to-go $ W_t^{\sigma^i,T} $ for any agent $ i $ and strategy $ \sigma^i $  as
\begin{equation}\label{eqr2gfh}
W_t^{\sigma^i,T}(z_{1:t},x_{1:t}^i) = \mE^{\sigma^i,\tsigma^{-i}} \big[ \sum_{n=t}^T \delta^{n-t} R(X_n^i,A_n^i,Z_n)
+ \delta^{T+1-t} G(Z_{T+1},X^i_{T+1}) \mid z_{1:t},x_{1:t}^i \big].
\end{equation}
Here agent $ i $'s strategy is $ \sigma^i $ whereas all other agents use strategy $ \tsigma^{-i} $ defined above. Since $ \mX,\mA $ are assumed to be finite and $ G $ absolutely bounded, the reward-to-go is finite $ \forall $ $ i,t,\sigma^i,z_{1:t},x_{1:t}^i $.
In the following, any quantity with a $T$ in the superscript refers the finite horizon model with terminal reward $G$.
\begin{lemma}\label{thmfh1}
	For any $ t \in [T] $, $ i \in \mN $, $ z_{1:t},x_{1:t}^i $ and $ \sigma^i $,	
	\begin{equation} \label{eqintlem}
	V_t^{T}(z_t,x_t^i) \ge \mE^{\sigma^i,\tsigma^{-i}} \big[ R(x_t^i,A_t^i,z_t)
	+ \delta V_{t+1}^{T}\big( \phi(z_t,\theta[z_t]) , X_{t+1}^i \big) \mid z_{1:t},x_{1:t}^i \big].
	\end{equation}
\end{lemma}

The result below shows that the value function from the backwards recursive algorithm is higher than any reward-to-go.

\begin{lemma}\label{thmfh2}
	For any $ t \in [T] $, $ i \in \mN $, $ z_{1:t},x_{1:t}^i $ and $ \sigma^i $,	
	\begin{gather}
	V_t^{T}(z_t,x_t^i) \ge W_t^{\sigma^i,T}(z_{1:t},x_{1:t}^i).
	\end{gather}
\end{lemma}
\proof
We use backward induction for this. At time $ T $, using the maximization property from~\eqref{eq:m_FP} (modified with terminal reward $ G $),
\begin{subequations}
	\begin{align}
	&V_T^{T}(z_T,x_T^i)
	\\
	&\triangleq \mE^{\tilde{\gamma}_T^{i,T}(\cdot \mid x_T^i),\tilde{\gamma}_T^{-i,T}} \big[ R(X_T^i,A_T^i,Z_T) + \delta G\big( \phi(z_T,\tilde{\gamma}_T^T)),X_{T+1}^i \big) \mid z_T,x_T^i \big]
	\\
	&\ge \mE^{{\gamma}_T^{i,T}(\cdot \mid x_T^i),\tilde{\gamma}_T^{-i,T}} \big[ R(X_T^i,A_T^i,Z_T)
	+ \delta G\big( \phi(z_T,\tilde{\gamma}_T^T)) ,X_{T+1}^i \big) \mid z_{1:T},x_{1:T}^i \big]
	\\
	&= W_T^{\sigma^i,T}(h_T^i)
	\end{align}
\end{subequations}
Here the second inequality follows from~\eqref{eq:m_FP} and~\eqref{eq:Vdef} and the final equality is by definition in~\eqref{eqr2gfh}.

Assume that the result holds for all $ n \in \{t+1,\ldots,T\} $, then at time $ t $ we have
\begin{subequations}
	\begin{align}
	&V_t^{T}(z_t,x_t^i)
	\\
	&\ge \mE^{\sigma_t^i,\tsigma_t^{-i}} \big[ R(X_t^i,A_t^i,Z_t)
	+ \delta V_{t+1}^{T}\big( \phi(z_t,\theta[z_t]) , X_{t+1}^i \big) \mid z_{1:t},x_{1:t}^i \big]
	\\
	&\ge \mE^{\sigma_t^i,\tsigma_t^{-i}} \big[ R(X_t^i,A_t^i,Z_t)
	+ \delta \mE^{\sigma^i_{t+1:T},\tsigma_{t+1:T}^{-i}} \big[ \sum_{n=t+1}^T \delta^{n-(t+1)} R(X_n^i,A_n^i,Z_n)
	\\ \nonumber
	&+ \delta^{T-t} G(Z_{T+1},X_{T+1}^i) \mid z_{1:t},x_{1:t}^i,Z_{t+1},X_{t+1}^i \big] \mid z_{1:t},x_{1:t}^i \big]
	\\
	&= \mE^{\sigma^i_{t:T},\tsigma^{-i}_{t:T}} \big[ \sum_{n=t}^T \delta^{n-t} R(X_n^i,A_n^i,Z_n)
	+ \delta^{T+1-t}G(Z_{T+1},X_{T+1}^i) \mid z_{1:t},x_{1:t}^i \big]
	\\
	&= W_t^{\sigma^i,T}(z_{1:t},x_{1:t}^i)
	\end{align}
\end{subequations}
Here the first inequality follows from Lemma~\ref{thmfh1}, the second inequality from the induction hypothesis, the third equality follows since the random variables on the right hand side do not depend on $\sigma_t^i$, and the final equality by definition~\eqref{eqr2gfh}.
\endproof

The following result highlights the similarities between the fixed-point equation in infinite-horizon and the backwards recursion in the finite-horizon.

\begin{lemma}\label{lemfhtoih}
	Consider the finite horizon game with $ G \equiv V $. Then $ V_t^{T} = V$,  $ \forall $ $ i \in \mN $, $ t \in \{1,\ldots,T\} $ satisfies the backwards recursive construction stated above (adapted from \eqref{eq:m_FP} and \eqref{eq:Vdef}).

\end{lemma}	
\proof
	Use backward induction for this. Consider the finite horizon algorithm at time $ t=T $, noting that $ V_{T+1}^{T} \equiv G \equiv  V $,
	\begin{subequations} \label{eqfhT}
		\begin{align} 	
		\tilde{\gamma}_T^{T}(\cdot \mid x_T^i) &\in \arg\max_{\gamma_T(\cdot \mid x_T^i)} \!\!\! \mE^{\gamma_T(\cdot \mid x_T^i)} \big[ R(x_T^i,A_T^i,z_T)
		+ \delta V\big( \phi(z_T,\tgamma_t^T) , X_{T+1}^i \big) \mid z_T,x_T^i \big]
		\\
		V_T^{T}(z_T,x_T^i) &= \mE^{\tilde{\gamma}_T^{T}(\cdot \mid x_T^i)} \big[ R(x_T^i,A_T^i,z_T)
		+ \delta V\big( \phi(z_T,\tgamma_t^T) , X_{T+1}^i \big) \mid z_T,x_T^i \big].
		\end{align}
	\end{subequations}
	Comparing the above set of equations with~\eqref{eq:m_FP_ih}, we can see that the pair $ (V,\tilde{\gamma}) $ arising out of~\eqref{eq:m_FP_ih} satisfies the above. Now assume that $ V_n^{T} \equiv V $ for all $ n \in \{t+1,\ldots,T\} $. At time $ t $, in the finite horizon construction from~\eqref{eq:m_FP},~\eqref{eq:Vdef}, substituting $ V$ in place of $ V_{t+1}^{T} $ from the induction hypothesis, we get the same set of equations as~\eqref{eqfhT}. Thus $ V_t^{T} \equiv V $ satisfies it.
\endproof

\section{}
\label{app:idih}
\proof
We prove this by contradiction. Suppose for the equilibrium generating function $\theta$ that generates MPE $\tsigma$, there exists $t\in[T], i\in[N], z_{1:t}\in\cH_t^c,$ such that \eqref{eq:m_FP} is not satisfied for $\theta$
i.e. for $\tgamma_t = \theta[z_t] = \tsigma(\cdot|z_t,\cdot)$,
\eq{
 \tilde{\gamma}_t \not\in \arg\max_{\gamma_t(\cdot|x_t)} \E^{\gamma_t(\cdot|x_t)} \left\{ R(X_t^i,A_t^i,Z_t) + \delta V(\phi(Z_t,\tilde{\gamma}_t), X_{t+1}) \big\lvert  x_t^i,z_t \right\} . \label{eq:FP4}
  }
  Let $t$ be the first instance in the backward recursion when this happens. This implies $\exists\ \hat{\gamma}_t$ such that
  \eq{
  \E^{\hat{\gamma}_t(\cdot|x_t)} \left\{ R(X_t^i,A_t^i,Z_t)+ \delta V(\phi(Z_t, \tilde{\gamma}_t), X_{t+1}) \big\lvert  z_{1:t},x_{1:t}^i\right\}
  \nn\\
  > \E^{\tgamma_t(\cdot|x_t)} \left\{ R(X_t^i,A_t^i,Z_t) +\delta V(\phi(Z_t, \tilde{\gamma}_t), X_{t+1}^i) \big\lvert  z_{1:t},x_{1:t}^i \right\} \label{eq:E1}
  }
  This implies for $\hat{\sigma}(\cdot|z_t,\cdot) = \hat{\gamma}_t$,
  \eq{
  &\E^{\tsigma} \left\{ \sum_{n=t}^{\infty} \delta^{n-t} R(X_n^i,A_n^i,Z_n) \big\lvert  z_{1:t-1}, x_{1:t}^i \right\}
  \nn\\
\\
  &= \E^{\tsigma_t^{i},\tsigma_t^{-i}} \left\{ R(X_t^i,A_t^i,Z_t) + \E^{\tsigma_{t+1:T}^{i} \tsigma_{t+1:T}^{-i}}   \left\{ \sum_{n=t+1}^{\infty} \delta^{n-t}R(X_n^i,A_n^i,Z_n) \big\lvert z_{1:t-1},Z_{t+1}, x_{1:t}^i,X_{t+1}^i \right\}  \big\vert z_{1:t}, x_{1:t}^i \right\} \label{eq:E2}
  \\
  &=\E^{\tgamma_t(\cdot|x_t) \tilde{\gamma}^{-i}_t} \left\{ R(X_t^i,A_t^i,Z_t) +\delta V (\phi(Z_t, \tilde{\gamma}_t), X_{t+1}^i) \big\lvert  z_t,x_t^i \right\} \label{eq:E3}
  \\
  &< \E^{\hat{\sigma}_t(\cdot|z_t,x_t^i) \tilde{\gamma}^{-i}_t} \left\{ R(X_t^i,A_t^i,Z_t) +\delta V(\phi(Z_t, \tilde{\gamma}_t), X_{t+1}^i) \big\lvert  z_t,x_t^i \right\}\label{eq:E4}
  \\
  &= \E^{\hat{\sigma}_t \tsigma_t^{-i}} \left\{ R(X_t^i,A_t^i,Z_t) +  \E^{\tsigma_{t+1:T}^{i} \tsigma_{t+1:T}^{-i}}\left\{ \sum_{n=t+1}^{\infty} \delta^{n-t}R(X_n^i,A_n^i,Z_n) \big\lvert z_{1:t},Z_{t+1}, x_{1:t}^i,X_{t+1}^i\right\} \big\vert z_{1:t}, x_{1:t}^i \right\}\label{eq:E5}
  \\
  &=\E^{\hat{\sigma}_t,\tsigma_{t+1:T}^{i} \tsigma_{t:T}^{-i}} \left\{ \sum_{n=t}^{\infty} \delta^{n-t} R(X_n^i,A_n^i,Z_n) \big\lvert  z_{1:t}, x_{1:t}^i \right\},\label{eq:E6}
  }
  where \eqref{eq:E3} follows from the definitions of $\tgamma_t$ and Appendix~\ref{app:D}, \eqref{eq:E4} follows from \eqref{eq:E1} and the definition of $\hat{\sigma}_t$, \eqref{eq:E5} follows from Appendix~\ref{app:D}. However, this leads to a contradiction since $\tsigma$ is an MPE of the game.
\endproof

\bibliographystyle{IEEEtran}

\medskip
\small
\end{document}